\documentclass[11pt,dvips]{article}

\usepackage{graphicx}
\usepackage{epsfig}

\usepackage{amssymb,amsfonts}       

\usepackage{mathrsfs}

\begin{document}

\def\uvel{\mathfrak{u}}
\def\onlinecite{\cite}
\def\eng{\varepsilon}
\def\Eng{{\mit {\sf  E }}}
\def\Efb{{\bf E}}
\def\g{\gamma}
\def\Ksub{{\mbox{\tiny${K}$}}}
\def\w{\omega{}}
\def\W{{\mit \Omega}}
\def\lam{\lambda}
\def\Lam{{\mit \Lambda}}
\def\th{\theta}
\def\dagsup{{\mbox{\tiny${\dagger}$}}}
\def\ddagsup{{\mbox{\tiny${\ddagger}$}}}
\def\aph{\alpha}
\def\D{\Delta}
\def\d{\delta}
\def\Taum{{\mit \Tau}}
\def\Nu{{\cal V}}
\def\hp{{h^*}}
\def\pd{\partial}
\def\lf{\left}
\def\rt{\right}
\def\NLamLw{N_{\Lam \varphi}}
\def\fsub{{_{^{f}}}}
\def\rhov{\rho_0}
\def\arm{{\rm a}}
\def\brm{{\rm b}}
\def\crm{{\rm c}}
\def\Tau{\Gamma{}}

\def\Bssub{{\mbox{\tiny${B}$}}}

\def\Vsup{{\mbox{\tiny${ \{V\}}$}}}
\def\osup{{\mbox{\tiny${ \{0\}}$}}}

      \def\Va{{\mathscr{V}}}




\title{{\LARGE\bf 
              Spectral Emission of Moving Atom 
 }
}
\bigskip\smallskip
\author{J X  Zheng-Johansson
\\ 
{\small\it   Institute of Fundamental Physics Research, Nyk\"oping, Sweden}
}
\maketitle
        \baselineskip 0.45cm
\begin{abstract}A renewed analysis of the H.E. Ives and G.R. Stilwell's experiment on moving hydrogen canal rays (J. Opt. Soc. Am., 1938, v.28, 215) concludes that the spectral emission of a moving atom exhibits always a redshift which informs not the direction of the atom's motion. The conclusion is also evident from a simple energy relation: atomic spectral radiation is emitted as an orbiting electron consumes a portion of its internal energy on transiting to a lower-energy state which however has in a moving atom  an additional energy gain; this results in a redshift in the emission frequency.  
Based on auxiliary experimental information and a scheme for de Broglie particle formation, we give a vigorous elucidation of the mechanism for deceleration radiation of atomic electron; the corresponding prediction of the redshift is in complete agreement with the Ives and Stilwell's experimental formula. 
\end{abstract}

\baselineskip 0.45cm

\section{Introduction}\label{Sec-intr}
Charged de Broglie particles such as the electron 
and the proton can be decelerated by emitting electromagnetic radiation. This occurs in all different kinds of processes, including atomic spectral emission produced in laboratory\cite{IvesStilwell1938,Kuhn:1962} and celestial processes\cite{Freedman99}, and charged particle synchrotron radiation   \cite{Crasemann98,WinickDoniach1980}.
The electromagnetic radiation emission from sources of this type is in common converted from a portion of the {\it internal energy} or the {\it mass} of a de Broglie particle involved, which often involves a final state in motion, hence moving source.  The associated source-motion effect has, except for admitting a relativistic effect connected to high sour-ce velocity, thus far been taken as no different from the ordinary Doppler effect that consists in a red- or blue- shift depending on the source is moving away or toward the observer. 
The ordinary Doppler effects are directly observable with  
moving sources of a "conventional type", like an external-field-driven oscillating electron, 
            %
             %
 an automobile horn, and others, that are externally driven into oscillation which does not add directly to the mass of the source. In this paper we first (Sec. \ref{Sec-2}) examine the property, prominently an invariable redshift, of moving atom radiation as informed by the  hydrogen canal ray experiment of Ives and Stilwell's \cite{IvesStilwell1938} performed at the Bell Labs in 1938 for a thorougher investigation of the associated anomalous Doppler effect then known.
Combining  with auxiliary experimental information and a  scheme for de Broglie particle formation\cite{Ref1}, 
we then elucidate (Secs. \ref{Sec-3}-\ref{Sec-5}) the mechanism for spectral emission of moving atom, or in essence the underlying (relative) deceleration radiation of moving de Broglie electron,  
and predict  Ives and Stilwell's experimental formula for redshift.

\section{Indication by Ives-Stilwell's experiment on \\ fast moving hydrogen atoms}\label{Sec-2}

\begin{figure}[] \begin{center} \leavevmode \hbox{\epsfxsize=10cm  \epsfbox{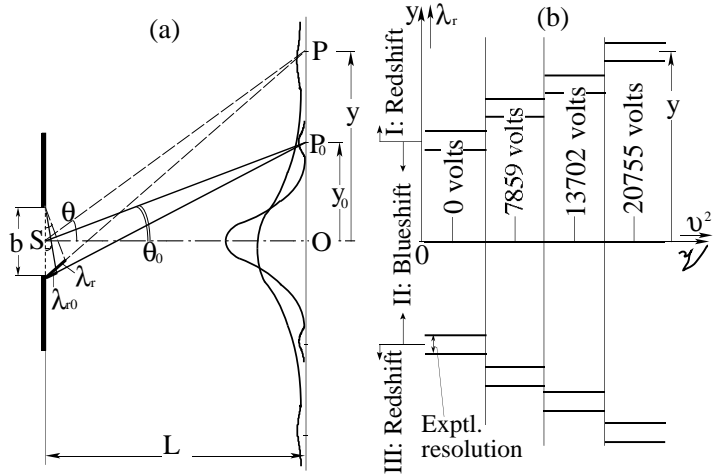}} 
          \vspace{-0.8cm}
\caption{\baselineskip 0.45 cm
(a) Schematic 
single-slit diffraction grating. (b) Experimental spectrogram, peak coordinates $y$($\propto \lam_r$) at several  voltages $\Va$ ($\propto v^2$), 7859, \ldots, 20755 volts,  after original Figure of Ref. \cite{IvesStilwell1938}. 
Spectral lines at finite  $\Va$ values all fall  in the redshift regions I and III beyond the $\Va=0$ ($v^2=0$)-lines illustrated in this plot. }\label{fig-fringe}
    \end{center}\leavevmode\vspace{-0.7cm}
\end{figure}

In their experiment on fast moving hydrogen canal ray spectral emission\cite{IvesStilwell1938},
Ives and Stilwell let positively charged hydrogen ions H$_i^+$ of mass $M_{H_i}$ and charge $q_i$ ($i=2,3$) be accelerated into a canal ray of high velocity, $v$, across accurately controlled electric potential $\Va$ 
correlated with $v$ 
 through the work-energy relation $q \Va = \frac{1}{2} M_{H_i} v^2$; 
or 
\begin{eqnarray}  \label{eq-v1}
v/c=A \sqrt{ \Va} 
\end{eqnarray}
with $c$ the speed of light, and $ A=  \sqrt{\frac{2 q_i }{c^2 M_{H_i}} }$. For $ \Va \sim 6700 \sim 20755 
 $ volts, $v \sim 
10^6$ m/s as from (\ref{eq-v1}).  
By neutralization and dissociation the ions are at exit converted to excited atoms that are unstable and will transit to ground state by emitting Balmer spectral lines. 
The wavelength, $\lam_r$, of the emitted H$\beta$ line is then measured using diffraction grating (Figure \ref{fig-fringe}a) as a function of $\Va$. For a finite $v$, the spectral line produces a first-diffraction peak at $P(v)$, at distance $y(v)=PO$ from the center $O$; for a hydrogen at rest, $v=0$, the line has a wavelength $\lam_{r0} =$ 4861 angst.  and produces a first peak at   $P_0$, $y_0=P_0O$. These have the geometric relations: $\lam_r=\frac{\lam_{r0}}{y_0} y$, and 
\begin{eqnarray} \label{eq-lamr1}
\D \lam_r=\lam_r-\lam'_{r0}= (\lam_{r0}/y_0) (y-y_0) \end{eqnarray} 
$\D \lam_r$ being the mean displacement of the               Doppler lines at a given $v$.
\input epsf   
  The measured spectrogram, Figure \ref{fig-fringe}b, informs  $y-y_0 = B' \sqrt{\Va}$ with $B'$ a constant; this combining with (\ref{eq-lamr1}) is:  
\begin{eqnarray}\label{eq-lamV}
\qquad 
\D \lam_r/\lam_{r0}
=(\lam_r-\lam'_{r0})/\lam_{r0}=B \sqrt{\Va} 
\end{eqnarray}
where $B=B'\lam_{r0}/y_0$.
If {\it assuming} 
$$\refstepcounter{equation}\label{eq-Dlamr}
\frac{\D \lam_r}{\lam_{r0}}= +\frac{v}{c}, 
\eqno(\ref{eq-Dlamr})
$$ 
then this and (\ref{eq-lamV}) give $\frac{v}{c}=  B \sqrt{\Va} $. But $\frac{v}{c} $ and $\sqrt{\Va}$ must satisfy (\ref{eq-v1}); thus $B\equiv A$; that is  
(\ref{eq-lamV}) writes: 
$$\D \lam_r/\lam_{r0} =A\sqrt{\Va}. \eqno(\ref{eq-lamV}') $$
 In \cite{IvesStilwell1938}, the two variables  $\frac{\D \lam_r}{\lam_{r0}}$ and $\sqrt{\Va}$ are separately measured and thus given an experimental relation, shown in Figure 10, of 
[1, p.222]
which agrees completely with  (\ref{eq-lamV})$'$; 
accordingly (\ref{eq-Dlamr}) is directly confirmed.
Furthermore there is a shift of center of gravity of  $\lam'_{r0}$ from $\lam_{r0}$: $\D' \lam_r=\lam_{r0}-\lam'_{r0}= \frac{1}{2}(\frac{v}{c})^2$; or $\lam'_{r0}= \lam_{r0} \lf(1-\frac{1}{2}(\frac{v}{c})^2\rt)\simeq \lam_{r0} \sqrt{1-(v/c)^2}$. With this and   (\ref{eq-Dlamr}) in $\lam_r=\lam'_{r0}+\D \lam_{r0}$ as given by the first equation of (\ref{eq-lamV}) or  of (\ref{eq-lamr1}), one gets:   
$$ \refstepcounter{equation}\label{eq-lampD}
\lam_r 
= \sqrt{1-(v/c)^2}  \lam_{r0}+ (v/c) \lam_{r0}.          
                     \eqno(\ref{eq-lampD})
$$ 
(\ref{eq-lampD}) gives 
$\lam_r -\lam_{r0}> \frac{v}{c}-\frac{1}{2}(\frac{v}{c})^2 \ge 0$; or,  $\lam_r $ is always elongated for $|v|>0$. Furthermore, (\ref{eq-Dlamr})--(\ref{eq-lampD}) are  obtained in \cite{IvesStilwell1938} for  both the cases where source and observer move toward  and away from each other: 
The  source velocity $v$ is  in the fixed $+x$-direction;
waves emitted parallel with  $v$ (Figure \ref{fig-2}) 
strike on the diffraction grating D (observer 1) directly (Figure \ref{fig-2}b), 
and waves antiparallel with $v$ (Figure \ref{fig-2}c)  strike on mirror M (observer 2) first and are then reflected to D. That is, (\ref{eq-lampD}) is regardless of the direction of the vector $c$.
 Therefore from Ives and Stilwell's experiment  we conclude:

\begin{center}
     \begin{minipage}[t]{13 cm}\baselineskip 0.45 cm
 The wavelength of spectral line emitted from an  atom in motion is always {\it longer},  or {\it red-shifted}, than from one at rest, irrespective if the atom is moving away or toward the observer; the faster the atom moves, the longer wavelength its spectral line is shifted to. 
\end{minipage}
\end{center}
\noindent
This apparently contrasts with the conventional Doppler effect where wavelengths will be  $ \lam_r= \lam_{r0}(1-v/c) $ and $\lam_r= \lam_{r0}(1+v/c)$ and show a blue or red shift according to if the source is moving toward or away from the observer.

\section{Emission frequency of a moving atom}\label{Sec-3}
If a H atom is at rest in the vacuum, its electron, of charge $-e$ in circular motion at  velocity $\uvel_{n+1}$ about the atomic  nucleus in an excited $n+1$th orbit, has from quantum-mechanical solution (and also solution based on the unification scheme \onlinecite{Ref1}) an eigen energy  
 $\eng_{u.n+1}=  -\hbar^2/[2   m_{e_0} (n+1)^2 {a_{\Bssub_0}}^2] $, $n=1, 2, \ldots$,
where $m_{e_0} =\g_0 M_e$, $\g_0=1/[1-(u_{n+1}/c)^2]^{-1/2}$ with $u_{n} (\sim 10^6$ m/s) being high, $M_e$  the electron rest  mass, and 
$a_{\Bssub_0}$ Bohr's radius (should already contain  
$1/\g_0$, see below).
If now the electron transits to an unoccupied $n$th orbit,  the atom lowers its energy to  $\eng_{u.n}$ and emits an electromagnetic 
wave of frequency
$$\refstepcounter{equation} \label{eq-canal-GLT0}
\hfill
 \nu_{r0} = \frac{\eng_{u.n+1}(0)-\eng_{u.n}(0)}{h} 
= \frac{ \hbar^2 (2n+1)}{h  2 m_e (n+1)^2n^2  a_\Bssub^2};   \hfill 
\eqno(\ref{eq-canal-GLT0})
$$
accordingly $ \lam_{r0} =c/\nu_{r0}$ and $ 
 k_{r0}= 2\pi/\lam_{r0} =2\pi \nu_{r0}/c $.

If now the atom is moving at a velocity $v$ in $+x$-direction,  $(v/c)^2>>0$, then in the motion direction, its orbital radius is  Lorentz  contracted to  $a_\Bssub= a_{\Bssub_0}/\g$, and its mass augmented according to  Einstein to $m_e=\g m_{e_0}= \g\g_0 M_e$ (see also the classical-mechanics solutions in \cite{Ref1}), where 
$\g=1/\sqrt{1-(v/c)^2}.$ With $ a_\Bssub$  and $m_e$ for $ a_{\Bssub_0}$  and $m_{e_0}$ in (\ref{eq-canal-GLT0}), we have $\nu_r=\frac{\eng_{u.n+1}(v)-\eng_{u.n}(v)}{h} =\g \nu_{r 0} $; including in this an additional term   $\delta \nu_r$ which we will justify below to  result because of an energy gain of the moving source, the spectral frequency for the $n+1$ $\rightarrow$ $n$ transition for the moving atom then writes
$$\refstepcounter{equation}\label{eq-canal-GLT0p}
  \nu_r= \g (\nu_{r 0} + \delta \nu_r). 
                  \eqno(\ref{eq-canal-GLT0p})
$$

\section{Atomic spectral emission scheme}
                  \label{Sec-AtmEmitSchm}\label{Sec-4}
\begin{figure}[here] \begin{center} \leavevmode \hbox{\epsfxsize= 10cm  \epsfbox{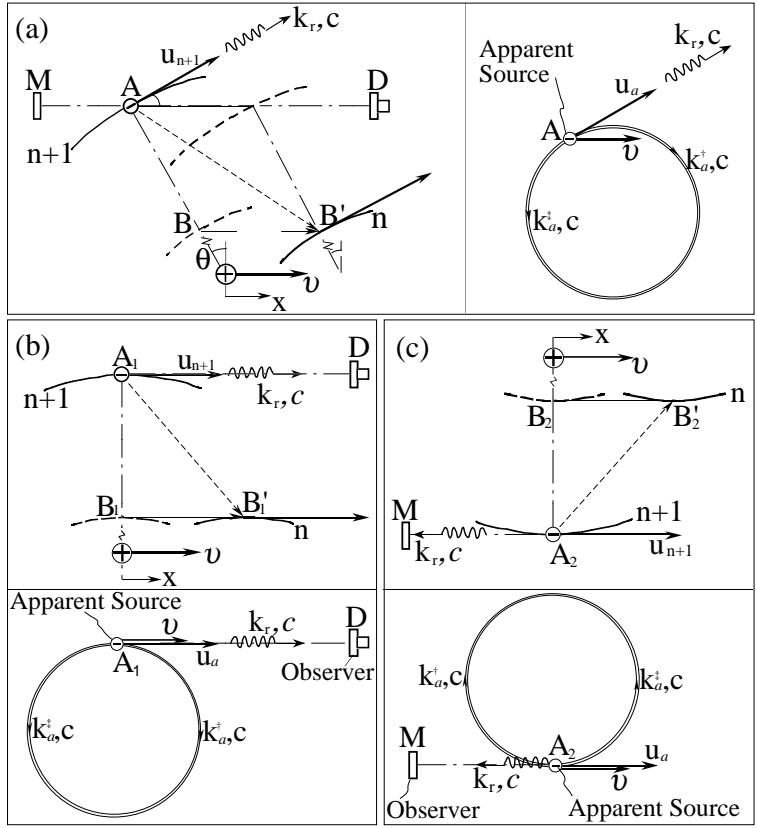}} 
\leavevmode
\vspace{-0.2cm}
\caption{\baselineskip 0.45 cm
An atomic electron comes off orbit $n+1$  statistically  e.g. at $A$ in (a), emitting in brief time $\delta t$ a single electromagnetic wave  of energy $h \nu_r $ in forward ($u_{n+1}$) direction, and then migrates  (transits) along  AB, $\perp  u_{n+1} $,  to orbit  $n$ for an atom of $v=0$, and across $AB'$ in time $ t_{AB'}$ for finite $v$ in $+x$-direction;  BB$' = v t_{AB'}$. In (a): $\angle{c,v}=\theta$; (b):  
  $c\| v$; (c): $-c \| v$. The insets in (a)-(c) illustrate the radiation from an apparent source.
} 
\label{fig-2}
\end{center}
\vspace{-0.5cm}
\end{figure}

We now inspect how an electron transits, from an initial $n+1$th to final $n$th orbit in an atom  moving in general, here at velocity $v$ in $+x$-direction. To the initial-state electron, with a velocity $u_{n+1}$ if $v=0$, the  finite ${ v}$ of the traveling atom will at each point on the orbit project a component $v \cos \theta $ onto $u_{n+1}(\th)$, with $\th$ in ($0,2\pi$); the average is 
 $\bar{u}_{n+1}=\int_{\th=0}^{2\pi}[{ u}_{n+1}+ v \cos \theta] d \th=u_{n+1}$.  That is, $\bar{u}_{n+1}$ and any its derivative dynamic quantities of the stationary-state orbiting electron are not affected by $v$ except through the second order factor $\g(v)$. The situation however differs during the $n+1\rightarrow n$ transition which distinct features may be induced as follows:  

(i) The transition ought  realistically be a mechanical process in which, in each sampling, the electron comes off orbit $n+1$ at a single definite location, e.g. $A$ in Figure \ref{fig-2}a. That where $A$ is located on the orbit in any sampling, is a statistic event.

(ii) The spectral radiation is a single monochromatic electromagnetic wave emitted in forward direction of the orbiting electron at the point ($A$) it comes off orbit $n+1$, as based on observations for decelerating electron radiation in a storage ring in synchrotron experiments \cite{Crasemann98}, which is no different from an orbiting atomic  electron except for its macroscopic orbital size.

(iii) It follows from (i)--(ii) combined with momentum conservation condition that the transition electron coming off at $A$, will migrate across shortest-distance $AB$, perpendicular to $u_{n+1}$, to orbit $n$, 
\input epsf  
at $B$ if the atom is at rest, or at $B'$ if the atom is moving at velocity $v$ in $x$-direction, given after vector addition.

(iv) A stationary-state orbiting electron on orbit $n^*$ ($=n+1$ or $n$), $\psi_{k_{dn^*}}$, is\cite{Ref1} a ({\it single}) beat or de Broglie phase wave convoluted from the opposite-traveling component total waves $\{\varphi^j_{kn}\}$ generated by an oscillatory massless  (vaculeon) charge $-e$, of wavevectors $k^{\dagsup}_{n^*}$ and $k^{\ddagsup}_{n^*}$, which being Doppler shifted for the  source moving at velocity $u_{n+1}^{j}$. An $n+1\rightarrow n$ transition emits  the difference between the two {\it single} waves, $\psi_{k_{dn+1}}$ and $\psi_{k_{dn}}$---the emitted radiation  is naturally also a {\it single} wave. And,

(v) The component total waves  making up the electron  beat wave  at $A$ is generated by the source in a brief time $\delta t$ when at $A$, and have a  wave  frequency $\sim \Nu= 511 $ keV$/h \simeq 10^{20}$ 1/s;  so the time for detaching the entire radiation wave train from the source is estimated $\delta t \sim 1/\Nu=8 \times 10^{-21} $ s. In contrast, the orbiting period of the electron is 
$ \tau_{d.n+1}=1/\nu_{d.n+1}=(n_1)^2 1.5 \cdot 10^{-16}$ s. 
So in time  $\delta t$, $<< \tau_{d.n+1}$, the electron is essentially not moved along orbit $n+1$ as well as path  $AB$ or $AB'$; hence $u_{n^*}$ ($\simeq u_{n+1}$) (thus $c$) and $v$ are at fixed angle $\th$.
Specifically if the electron comes off at $A_1$ and $A_2$
as in  Figure \ref{fig-2}b and c, respectively, we have the cases of the source and observer  approaching each other  and  receding from each other  
$$ \refstepcounter{equation}\label{eq-uv}
c \| v  \qquad {\rm and} \qquad 
-c \| v  
\eqno(\ref{eq-uv})
$$

The wave and dynamic variables for the nonstationary transition process would not be a simple difference between the solutions for the stationary statesm. However, we can try to represent the process  effectively using an apparent source such that

(v.1) the total wave detached from the apparent sour-ce gives  the same observed radiation as due to the actual source; and 

(v.2) the apparent source in transition has  the same motion as the (actual source of the) transition electron, that is, translating at the velocity  $v$ (cf. item iv) in $+x$-direction here.

\section{A theoretical formula for the redshift}
                            \label{Sec-EqRS} \label{Sec-5}

In fulfilling (v.1), the apparent source ought to be an oscillatory charge ($q$) executing in stationary state circular motion at velocity $u_{a} $ on orbit $n+1$ (insets in Figure \ref{fig-2}). Let first the orbit $n+1$ be at rest, $v=0$, and so must be the apparent source  as by (v.2).
The apparent source generates two identical monochromatic electromagnetic waves traveling oppositely along orbit $n+1$, of wavevectors $ k_{a0}^{\dagsup} = k_{a0}^{\ddagsup}=k_{a0}$, which  superpose into a single electromagnetic wave $\psi_{k_{a0}}$. On transition, the source emits the entire $\psi_{k_{a0}}$ in the direction parallel with $u_a (\theta)$, by simply detaching it; thus  
 $k_{a0}\equiv k_{r0} =2\pi/\lam_{r0}$. 

Let now orbit $n+1$ be in motion at  velocity $v$ in $+x$-direction, and so must be the apparent source. Let the source comes off orbit $n+1$ at point $A_1$ (Figure \ref{fig-2}b). In a brief time $\delta t $ before this, the apparent source was essentially at $A_1$ and generating two waves $\varphi^{\dagsup}_{k_a^{\dagsup}}$ parallel and antiparallel with $u_{{a}}$, thus $v $; their wavelengths were owing to the source motion of $v$ Doppler shifted, to
         ${\lam_a^{\dagsup}}= \lam_{r0}(1-\frac{v}{c})$, ${\lam_a^{\ddagsup}}= \lam_{r0}(1+\frac{v}{c})$, and  wavevectors  
${k_a^{\dagsup}}= \frac{2\pi}{{\lam_{a}^{\dagsup}} }
$, ${k_a^{\ddagsup}}= \frac{2\pi}{{\lam_{a}^{\ddagsup}} }
$ with the  Doppler shifts 
$$\refstepcounter{equation}\label{eq-cank2}
\hfill  {k_a^{\dagsup}{}} -k_{r0} = \frac{(v/c)k_{r0} }{ 1-v/c} \quad(a); \qquad  
 \hfill  k_{r0}-{k_a^{\ddagsup}{}}= \frac{(v/c)k_{r0} }{ 1+v/c} \quad (b)
\hfill \eqno(\ref{eq-cank2})
$$
The two waves superpose to $\psi_{k_a}=\varphi^{\dagsup}_{k_a^{\dagsup}}+\varphi^{\ddagsup}_{k_a^{\ddagsup}}$, 
being according to \cite{Ref1} now a single beat, or de Broglie phase wave of the moving apparent source. On transition the source detaches the entire single beat wave $\psi_{k_a }$, which is no longer "regulated" by the source and will relax into a   pure electromagnetic wave $\psi_{k_r}$, but 
in conserving momentum, 
retains in the single direction  parallel with $u_a$ thus $v$. Similarly, if the source exits at $A_2$ (Figure \ref{fig-2}c),  a single electromagnetic  wave will be  emitted parallel with  $u_{{a}}(A_2)$, or, $-v$.    
$\psi_{k_a}$ has a de Broglie wavevector given\cite{Ref1} by the geometric mean of (\ref{eq-cank2}a) and (b):
\begin{eqnarray}
\label{eq-cank3}
 & k_{{a}.d} = \sqrt{
( {k_a^{\dagsup}} -k_{r0} )(k_{r0}-{k_a^{\ddagsup}})
 } 
=\frac{\lf(\frac{v}{c}\rt)k_{r0} }{\sqrt{1-(v/c)^2}} 
\cr
{\rm or} & \qquad k_{{a}.d} = \g \lf(\frac{v}{c}\rt)  k_{r0}.  \qquad \qquad\quad
\end{eqnarray}
We below aim to express the $k_{{a}.d} $-effected radiation variables $k_r$, $\nu_r$ and $\lam_r$,  which being directly observable. Momentum conservation requires $|\hbar k_{{a}.d}$ $|$ $=|\hbar \d k_r|$. 
$k_{{a}.d} $ is associated with an energy {\it gain}  of the apparent source, $\eng_{{a}.v}$ ($=\frac{(\hbar k_{a.d})^2}{2m_e}$), owing to its motion, and thus an energy  {\it deficit} in the emitted radiation wave $\psi_{k_r}$, 
\begin{eqnarray}
\d \eng_{r}(=\hbar \d k_r c)=-\eng_{{a}.v},   \nonumber
\\ \label{eq-caneng}
{\rm for  \ either  \ } c \|v \  {\rm or}  \  -c\| v.           
\end{eqnarray}
and accordingly momentum and frequency deficits in the emission
\begin{eqnarray}
\label{eq-cank3p}
& \d k_r = -k_{{a}.d} =-(v/c) k_{r0},  
\\
\label{eq-X1a}
&\d \nu_r = \d k_r  c 
= -(v/c) k_{r0}c
=-(v/c)\nu_{r0}.
   \end{eqnarray}
With (\ref{eq-X1a}) in (\ref{eq-canal-GLT0p}), we have 
\begin{eqnarray}   \label{eq-X3}
\nu_r 
= \g\lf(1 -\frac{v}{c}\rt)\nu_{r0} 
\simeq \g \nu_{r0}- \lf(\frac{v}{c}\rt) \nu_{r0}
          \end{eqnarray} 
where $\g $ in front of $\delta \nu_r $ is higher order and thus dropped. With (\ref{eq-X3}) we can further compute for the emitted wave:
\begin{eqnarray}
\label{eq-X4}
k_{r}= \frac{2\pi \nu_r}{c}
= \g\lf(1-\frac{v}{c}\rt)k_{r0} 
\simeq  \g k_{r0} -\lf(\frac{v}{c}\rt)k_{r0},
\\
\label{eq-X6}
\lam_r=\frac{c}{\nu_r}
= \frac{c}{\nu_{r0}(1/\g-v/c)}
\simeq \frac{\lam_{r0}}{\g}+\lf(\frac{v}{c}\rt)\lam_{r0}.
\end{eqnarray}
The theoretical prediction (\ref{eq-X6}) for $\lam_r$ above is seen to agree exactly with Ives and Stilwell's experimental formula, (\ref{eq-lampD}). Notice especially that the prediction gives  $\delta \nu_r <0$ and $\delta \lam_r> 0$ for both $c\| v$ and $-c \| v$ as follows from (\ref{eq-caneng}); that is, they represent always a redshift in the emission spectral line, regardless if the wave is emitted parallel or antiparallel with $v$. 

\section{Discussion}
From  the forgoing analysis of the direct experimental spectral data of Ives and Stilwell on hydrogen canal rays, and with the elucidation of the underlying mechanism, we conclude without ambiguity that, the spectral emission of a moving hydrogen atom exhibits always a redshift compared to that from an atom at rest; the faster the atom moves, the redder-shift it shows. This is not an ordinary Doppler effect associated with a conventional moving source, but rather is an energy deficiency resulting from the de Broglie electron kinetic energy gain in transition to a moving frame, a common feature elucidated in \cite{Unif2} to be exhibited by the deceleration radiation of all de Broglie particles.  This redshift does not inform the direction of  motion of the source (the atom).

It is on the other hand possible for an atomic spectral  emission to exhibit blue shift for other reasons, for example, when the observer is moving toward the source as based on Galilean transformation.  The author thanks P-I Johansson for his support of the research and the Studsvik library  for helping acquiring needed literature. 

\nopagebreak

\end{document}